\documentclass[aps,preprint,pra,,showpacs]{revtex4-1}
\usepackage{graphicx}

\begin{document}

\title{GHZ states and PR boxes in the classical limit}

\author{Daniel Rohrlich}
\affiliation{Department of Physics, Ben-Gurion University of the Negev, Beersheba
84105 Israel}

\author{Guy Hetzroni}
\affiliation{Program in the History and Philosophy of Science, The Hebrew University of Jerusalem, Jerusalem 91905, Israel}

\date{\today}

\begin{abstract}
A recent paper \cite{max80} argues that bipartite ``PR-box" correlations, though designed to respect relativistic causality, in fact $violate$ relativistic causality in the classical limit.  As a test of Ref. \cite{max80}, we consider GHZ correlations as a tripartite version of PR-box correlations, and ask whether the arguments of Ref. \cite{max80} extend to GHZ correlations.  If they do---i.e. if they show that GHZ correlations violate relativistic causality in the classical limit---then Ref. \cite{max80} must be incorrect, since GHZ correlations are quantum correlations and respect relativistic causality.  But the arguments fail.  We also show that both PR-box correlations and GHZ correlations can be retrocausal, but the retrocausality of PR-box correlations leads to self-contradictory causal loops, while the retrocausality of GHZ correlations does not.
\end{abstract}


\maketitle

Quantum mechanics might make more sense to us if we could derive it from simple axioms with clear physical content, instead of opaque axioms about Hilbert space.  Aharonov \cite{a} and, independently, Shimony \cite{s} conjectured that quantum mechanics might follow from the two axioms of nonlocality and relativistic causality (no superluminal signalling).  For example, quantum correlations respect relativistic causality, but they are nonlocal:  they violate the Bell-CHSH \cite{bell,bell1976local} inequality.  Could quantum mechanics be $unique$ in reconciling these axioms, just as the special theory of relativity is unique in reconciling the axioms of relativistic cauasality and the equivalence of inertial frames?  So-called ``PR-box" \cite{PR} correlations disprove this conjecture.  Like quantum correlations, they respect relativistic causality; but unlike quantum correlations, they violate the Bell-CHSH inequality $maximally$.  Nevertheless, a recent paper \cite{max80} argues that the addition of one minimal axiom of clear physical content---namely, the existence of a classical limit---suffices for ruling out PR-box correlations.

The additional axiom is minimal in the following sense:  Quantum mechanics has a classical limit in which there are no uncertainty relations; there are only jointly measurable macroscopic observables.  This classical limit---our direct experience---is an inherent constraint, a boundary condition, on quantum mechanics and on any generalization of quantum mechanics.  Thus PR-box correlations, too, must have a classical limit.  Reference \cite{max80} argues that in this classical limit, PR-box correlations (and, by extension \cite{max}, all stronger-than-quantum bipartite correlations) allow observers ``Alice" and ``Bob" to exchange superluminal signals.  The argument \cite{max80, max} relies on measurement sequences that are observable but exponentially improbable.  It is therefore of interest to test the argument by applying it to a different problem.  In particular, GHZ correlations \cite{GHZ} are a tripartite version of PR-box correlations in the sense of being all-or-nothing correlations (perfect correlations and anticorrelations).  Could Alice, Bob and an additional observer, ``Jim", use GHZ correlations, in the classical limit, to exchange superluminal signals?  Does the argument of Ref. \cite{max80} lead to this conclusion?  If so, it is clearly an incorrect argument:  quantum mechanics and its classical limit do $not$ violate relativistic causality!  The first section of this paper reviews the arguments of Ref. \cite{max80} and attempts to extend them to show how Alice, Bob and Jim could exchange superluminal signals in the classical limit; but these attempts fail.  The second section compares PR-box and GHZ correlations to show how retrocausality is self-contradictory in the first case but not in the second.

\section{GHZ correlations in the classical limit}
\label{GHZsignal}

Let Alice and Bob make spacelike separated measurements on pairs of particles.  For each pair (indexed by $i$), one member is in Alice's laboratory, and she can choose to measure observables $a_i$ or $a_i'$ (but not both) on it; the other member is in Bob's laboratory, and Bob can choose to measure observables $b_i$ or $b_i'$ (but not both) on it.  All four observables $a_i$, $a_i'$, $b_i$ and $b_i'$ take values $\pm 1$ with equal probability.  The definition of PR-box correlations,
\begin{equation}
	\label{PRboxCorr}
	C(a_i,b_i)=  C(a_i,b_i')= C(a_i',b_i)= 1=- C(a_i',b_i')~~~,
\end{equation}
implies that Alice can manipulate the correlations between the observables $b_i$, $b_i'$ of Bob's particle by choosing whether to measure $a_i$ or $a_i'$:  indeed, $b_i$ and $b_i'$ are perfectly correlated if she measures $a_i$ (as both of them are perfectly correlated with her outcome), and perfectly anticorrelated if she measured $a_i'$ (as $b_i$ is correlated with her outcome and $b_i'$ is anticorrelated with it).  Thus, even though Alice's choice of measurement does not affect Bob's distribution of either $b_i$ or $b_i'$, it does affect correlations between these two incompatible observables.  So can Alice exploit these correlations to signal to Bob?  No, she cannot, since Bob (by assumption) cannot measure both $b_i$ and $b_i'$.  Yet Ref.  \cite{max80} claims she can, {\it if} we assume a classical limit (in which the averages over large ensembles of complementary observables can be measured together).

To see how, let us consider an ensemble of $N$ pairs shared by Alice and Bob and obeying Eq. (\ref{PRboxCorr}).  Apparently, the $N$ pairs are just as useless for signalling as one pair, since, for each pair, Bob is allowed to measure only $b_i$ $or$ $b_i'$. But the classical limit assumption means that Bob can obtain $some$ information about such sums as
$B=\sum_{i=1}^N{b_i}/N$ and $B'=\sum_{i=1}^N{b'_i}/N$, because these can take macroscopic values $\pm 1$.  In the classical limit, all observables commute, so Bob must be able to obtain at least $some$ information both about $B$ $and$ about $B'$.

Now let us imagine two possible scenarios.  In one scenario, Alice measures $a_i$ consistently on all her $N$ particles.  In the other scenario, she measures $a'_i$ consistently on all her $N$ particles.  What does Bob obtain from his measurements?  The average value of $B$ is $\langle B\rangle =0$.  Even typical deviations of $B$ are small, i.e. of order $1/\sqrt{N}$, so they disappear in the classical limit.  Apparently the scenarios lead to the exact same conclusion:  Bob cannot read Alice's 1-bit message, encoded in her choice of what to measure.

Yet it will sometimes happen (with probability $2^{-N}$) that $B$ will take the value 1.  If Alice and Bob repeat either scenario exponentially many times, they can produce arbitrarily many cases of $B=1$.  True, there will be measurement errors in Bob's results, but in the classical limit Bob must obtain at least $some$ information about $both$ $B$ $and$ $B'$.  Now if Alice consistently measures $a_i$,  Bob can expect to obtain $B=1$ with probability close to $2^{-N}$.  And he can also expect to obtain $B=1=B'$ with the $same$ probability, and not with probability $2^{-2N}$, because Alice's choice has correlated $\langle B\rangle$ with $\langle B'\rangle$.  Conversely, if Alice consistently measures $a'_j$, then Bob can expect to obtain $B=1$ with probability close to $2^{-N}$, and he can also expect to obtain $B=1=-B'$ with the $same$ probability, and not with probability $2^{-2N}$, because Alice's choice has $anti$correlated $\langle B\rangle$ with $\langle B'\rangle$.  Another way for Bob to get Alice's message is to observe the variance in his measurements of $B\pm B'$: if Alice measures $a_i$ consistently, the distribution of $B+B'$ (over repeated trials with $N$ pairs at a time) is binomial, while the distribution of $B-B'$ has zero variance, and vice versa in the other scenario.  Thus Alice can send Bob a (superluminal) message in the classical limit.

It does not matter that the price of a one-bit message from Alice to Bob may be astronomical. As long as it is possible, at any price, it constitutes a violation of relativistic causality, which we cannot allow.  Hence PR-box correlations violate relativistic causality in the classical limit, as claimed.

Before proceeding to tripartite (GHZ) correlations, let us stop to consider bipartite $quantum$ correlations.  Does the above argument imply that they, too, allow signalling in the classical limit?  If so, it cannot be correct.  Most similar to PR-box correlations are quantum correlations that saturate Tsirelson's bound \cite{tsir} for the Bell-CHSH inequalities. Without loss of generality, we can consider entangled pairs of spin-1/2 particles in the state $\left[\vert \uparrow \rangle_{A} \vert \uparrow \rangle_{B} + \vert \downarrow \rangle_{A} \vert \downarrow \rangle_{B}\right]/\sqrt{2}$. In this state, Alice and Bob always obtain perfect correlations if they measure spin along the same axes in the $xz$ plane.

Quantum correlations saturate Tsirelson's bound when $a=\sigma_z^{A}$, $a'= \sigma_x^{A}$, $b= (\sigma_z^{B} +\sigma_x^{B} )/\sqrt{2}$ and $b'= (\sigma_z^{B} -\sigma_x^{B} )/\sqrt{2}$, where each of the four observables takes the values $\pm 1$.  (We suppress the index $i$.) Their correlations are
\begin{equation}
	\label{CHSHcorr}
	C(a,b)=C(a,b')= C(a',b)= \frac{\sqrt{2}}{2}=-C(a',b')~~~~.
\end{equation}
If Alice measures $a$, then $b$ and $b'$ are correlated with her results.  If she measures $a'$, then $b$ is correlated with her results and $b'$ is anticorrelated.  As in the discussion of PR-box correlations, we can compute and compare the variances of $(b+b')/\sqrt {2}$ vs. $(b - b')/\sqrt{2}$.  But, by definition, these observables correspond to $\sigma_z^{B}$ and  $\sigma_x^{B}$, respectively, i.e. to $a$ and $a'$ on Bob's particle in the pair, which is left in the same state as Alice's.  Now if Alice measures $a$ consistently on her particles and Bob measures $(b+b')/\sqrt {2}$, the variance in Bob's results is maximal just because the variance in Alice's results is maximal.  (That is, she has equal probability to obtain $\pm 1$.)  Conversely, if Alice measures $a$ consistently on her particles and Bob measures $(b-b')/\sqrt {2}$, the variance in Bob's results is maximal simply because a measurement of $\sigma_x^{B}$ after Alice measures $a$ is equally likely to be $\pm 1$, whatever Alice obtains.  We thus find that the correlations in Eq. (\ref{CHSHcorr}) are not strong enough to induce any difference between the variances of the observables $B+B'$ and $B-B'$.  Indeed, they are the strongest correlations that do not induce such a difference and therefor do not permit signalling in the classical limit \cite{max}.

Reference \cite{max80} claims that correlations that are too strong violate relativistic causality in the classical limit, and that PR-box correlations are too strong because they provide absolute ``all or nothing" correlations.  But quantum mechanics, as well, provides ``all or nothing" correlations.  Consider a triplet of spin-half particles in a GHZ state $\vert\Psi_{GHZ}\rangle=\left[ \vert \uparrow\rangle_{A} \vert \uparrow\rangle_{B} \vert
\uparrow\rangle_{J} - \vert \downarrow\rangle_{A}\vert \downarrow\rangle_{B}\vert
\downarrow\rangle_{J} \right]/\sqrt{2}$ shared by Alice, Bob and Jim in their respective laboratories.  Suppose that these observers measure either $\sigma_x$ or $\sigma_y$ on their respective particles. Let $a_x$ denote Alice's outcome from a measurement of $\sigma_x^{A}$ (the $x$ component of the spin of her particle) and let $a_y$ denote Alice's outcome from a measurement of $\sigma_y^{A}$ (the $y$ component of the spin), with analogous notations for Bob and Jim.  The state $\vert\Psi_{GHZ}\rangle$ is an eigenstate of the following four operators, satisfying
\begin{eqnarray}
	\label{GHZcorr}
\vert\Psi_{GHZ}\rangle
	 &=& \sigma_y^{A}\sigma_x^{B}\sigma_y^{J}\vert\Psi_{GHZ}\rangle\cr &=& \sigma_y^{A}\sigma_y^{B}\sigma_x^{J}\vert\Psi_{GHZ}\rangle\cr &=&
\sigma_x^{A}\sigma_y^{B}\sigma_y^{J}\vert\Psi_{GHZ}\rangle\cr
&=& -\sigma_x^{A}\sigma_x^{B}\sigma_x^{J}\vert\Psi_{GHZ}\rangle~~~~.
\end{eqnarray}
The implication is that if all three observers measure $\sigma_x$ on their particles, they will discover that $a_x b_x j_x=-1$. Similarly, if the appropriate measurements are carried out, they will discover that $a_x b_y j_y=1=a_y b_x j_y=a_y b_y j_x$ as in Eq. (\ref{GHZcorr}). In their famous paper \cite{GHZ}, Greenberger, Horne and Zeilinger (GHZ) used these facts to show that there is no way to assign simultaneous values consistently to all six variables $a_x$, $a_y$, $b_x$, $b_y$, $j_x$ and $j_y$.  This fact rules out any local hidden variable model for the GHZ state.

Can Alice, Bob and Jim use GHZ states to signal?  For definiteness, let us assume that Jim tries to send a signal to Alice and Bob via his choice of what to measure, $\sigma_x^{J}$ or $\sigma_y^{J}$.  Before going to the classical limit, let's ask whether Jim can send Alice and Bob a signal using just a few triplets.  Note that if Jim measures $\sigma_x^{J}$ and gets $j_x =-1$, then $a_x$ and $b_x$ must be correlated; we write $a_x b_x=1$.  In the same notation, $a_y b_y=-1$.  In fact, if Jim measures $\sigma_x^{J}$, we find $a_x b_x=-a_y b_y=1$ whatever he gets.  On the other hand, if Jim measures $\sigma_y^{J}$, we obtain the analogous equation $a_xb_y = a_yb_x$, whatever he gets, and no correlation between $a_x$ and $b_x$ or $a_y$ and $b_y$.  Are these correlations of any use?  Alice and Bob $cannot$ measure all their observables $a_x, a_y, b_x, b_y$ to infer Jim's choice.

But the commutation relations
\begin{equation}
	\label{GHZcomm}
[~\sigma_x^{A}\sigma_x^{B},~ \sigma_y^{A}\sigma_y^{B}~]=0=[~\sigma_x^{A}\sigma_y^{B},~ \sigma_y^{A}\sigma_x^{B}~]~~~,
\end{equation}
imply that Alice and Bob $can$ obtain $a_x b_x$ and $a_yb_y$ to see if they are anticorrelated or, alternatively, can obtain $a_x b_y$ and $a_y b_x$ to see if they are correlated!  In the first case, Jim must have measured $\sigma_x^{J}$ and in the second case, he must have measured $\sigma_y^{J}$.  Right?

Wrong.  This scheme fails.  To see why, we first note that if Alice and Bob measure both $\sigma_x^{A}\sigma_x^{B}$ and $\sigma_y^{A}\sigma_y^{B}$, they will certainly find that $a_xb_x = -a_yb_y$ simply because the product of operators $\sigma_x^{A}\sigma_x^{B} \sigma_y^{A} \sigma_y^{B}$ equals $-\sigma_z^A \sigma_z^B$ which yields $-1$ when applied to $\vert\Psi_{GHZ}\rangle$. Likewise, if Alice and Bob measure both $\sigma_x^{A} \sigma_y^{B}$ and $\sigma_y^{A}\sigma_x^{B}$, they will verify that $a_x b_y = a_y b_x$, simply because the product of operators $\sigma_x^{A}\sigma_y^{B} \sigma_y^{A} \sigma_x^{B}$ equals $\sigma_z^A \sigma_z^B$ which yields $1$ when applied to $\vert\Psi_{GHZ}\rangle$.  In fact, Alice and Bob can learn nothing about Jim's choice from their measurements.

We are back to square one.  So let us try to apply the classical-limit argument of Ref. \cite{max80}.  By analogy with Ref. \cite{max80}, let  Alice, Bob and Jim make collective measurements on ensembles of $N$ triplets at a time, with Jim measuring either $\sigma_x^{J}$ or $\sigma_y^{J}$ consistently on his particles.  For large enough $N$, we can define a collective variable $J_x = \sum{j_x}/N$, if Jim chooses to measure $\sigma_x^{J}$, or alternatively $J_y =\sum{j_y}/N$, if he chooses to measure $\sigma_y^{J}$, where the $j_x$ and $j_y$ represent Jim's particles in any given ensemble.  (As before, we suppress the index $i$.) We can then define also the collective variables $A_x =\sum{a_x}/N$, $A_y =\sum{a_y}/N$, $B_x =\sum{b_x}/N$ and $B_y =\sum{b_y}/N$.  In some (rare) cases, one or more of these collective variables may even reach $\pm 1$.  Above we noted that, for a given triplet of particles, Alice and Bob cannot measure all their observables $a_x$,  $a_y$, $b_x$ and $b_y$ to infer Jim's choice.  But, according to the classical-limit argument, there cannot be such complementary between $A_x$ and $A_y$, or between $B_x$ and $B_y$.  Alice and Bob must have access to at least $some$ information about all these variables.  True, their expectation values all vanish, but if Alice, Bob and Jim repeat their measurements exponentially many times, they will find fluctuations as large as $\pm 1$.  Since Eq. (\ref{GHZcorr}) involves products, we cannot directly sum over it to get a relation between $A_x$ or $A_y$ and $B_x$, $B_y$, $J_x$ and $J_y$.  Even so, suppose Jim measures $\sigma_x^{J}$ and obtains $j_x=-1$ for every particle in his ensemble.  Then for each of the other two particles in the triplet, $a_x$ and $b_x$ are correlated and $a_y$ and $b_y$ are anticorrelated.  But Alice and Bob will not be able to detect this correlation unless another ``miracle" occurs, in addition to the ``miracle" that happened in Jim's laboratory.  For example, suppose that $A_x =1$.  It follows from Eq. (\ref{GHZcorr}) that $B_x =1$ (up to fluctuations due to measurement errors).  Then Alice and Bob could compare their results for $A_x$ and $B_x$ to uncover a striking correlation between them and conclude that Jim had measured $J_x$ and not $J_y$.

But this conclusion can be valid only if the statistics support it.  In this scenario, we have assumed rare fluctuations:  $J_x =-1$ and $A_x=1$.  Since the two fluctuations are independent, their combined probability is the product of their individual probabilities, namely $2^{-N} \times 2^{-N} = 2^{-2N}$.  For this rare scenario, we don't need to assume also that $B_x =1$; Eq. (\ref{GHZcorr}) requires it.  Thus, with probabilty $2^{-2N}$, Alice and Bob will obtain $A_x=1=B_x$.  Does this result imply that Jim consistently measured $\sigma_x^{J}$ on his particles?  How likely is it that Alice and Bob would have obtained $A_x=1$ and $B_x=1$ if Jim had chosen to measure $\sigma_y^{J}$ on all his particles, making $a_x$ and $b_x$ uncorrelated?  The probability would have been $2^{-2N}$, exactly the same.  So, once again, Alice and Bob have no way of reading Jim's one-bit message (his choice of what to measure).

The statistics don't work out in the case of GHZ triplets as they do in the case of PR-box pairs.  We therefore conclude that despite the similarity between Eq. (\ref{PRboxCorr}) and Eq. (\ref{GHZcorr}), GHZ correlations do not allow Jim to signal to Alice and Bob by choosing which observable to measure (at least via the above attempts), even if we assume a classical limit in which they can measure the ensemble averages of incompatible observables.  The argument of Ref. \cite{max80} passes the test we prepared for it.

\section{Retrocausality in PR-box and GHZ correlations}
\label{causality}

Instantaneous signalling directly violates relativity theory, opening the door to causal loops and contradictions. In particular, consider the classical limit of a PR-box ensemble, with Alice sending one bit of information $i_A \in \{0,1\}$ to distant Bob.  In an ``unprimed" reference frame, Bob receives Alice's message instantaneously (at time $t_B=t_A$); but in an appropriate ``primed" reference frame, Alice's bit could be a message into the past, e.g. Bob receives her bit (at time $t_B^\prime$) before she sends it (at time $t_A^\prime>t_B^\prime$). Applying the principle of relativity, we infer that in the primed reference frame, Bob could send a bit $i_B\in \{0,1\}$ at time $t_B^\prime$ that Alice would receive instantaneously (at time $t_B^\prime$) before sending $i_A$. Then if Alice's device is set to echo whatever message she receives from Bob (so that $i_A=i_B$), and Bob's device is set to yield the inverse of the message he receives from Alice (so that $i_B=1-i_A$), together they create a self-contradictory causal loop, as in Fig. 1.

From this example it may seem obvious that PR-box correlations and GHZ correlations are distinguished, in that PR-box correlations in the classical limit can be retrocausal, and create self-contradictory causal loops, whereas GHZ correlations cannot be retrocausal.  It is therefore of interest to note that this distinction is $not$ valid.  GHZ correlations can be understood as retrocausal, as well! Yet the predictions implied by Eq. (\ref{GHZcorr}) do not create causal loops. How can quantum correlations affect distant or past events without creating causal loops?
\begin{figure}
\centerline{
\includegraphics*[width=80mm]{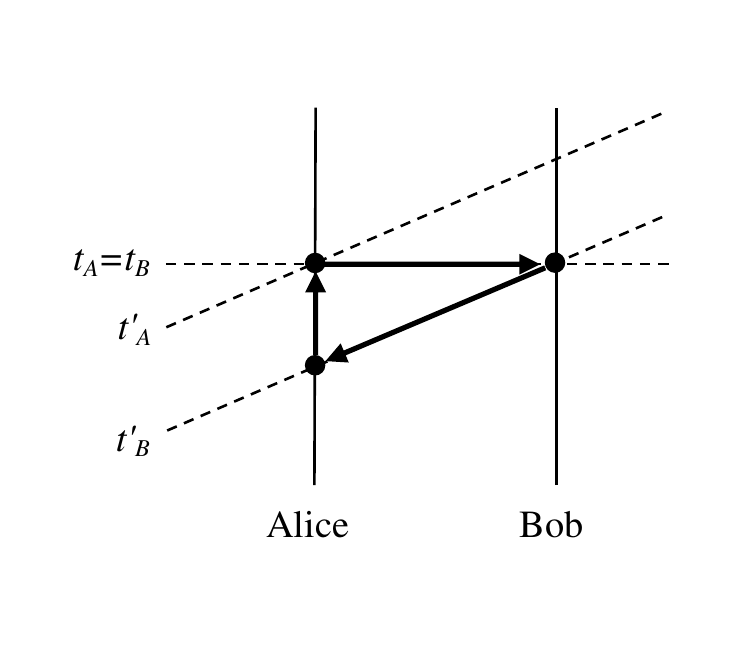}}
\caption[]{The horizontal dotted line represents an equal-time surface in the unprimed frame, while the tilted dotted lines represent two equal-time surfaces in the primed frame.  The arrows, each representing a cause and an effect, form a closed causal loop.}
\label{Fig1}
\end{figure}

Reference \cite{grunhaus1996jamming} imagines an action called ``jamming" in which Jim ``the Jammer" can, by pushing a button on a device he holds, decide at any moment whether to turn an ensemble of entangled pairs of particles shared by Alice and Bob into a product state.  Although jamming is action at a distance, it is consistent with relativistic causality if two conditions are met.  The first condition, the $unary$ condition, states that Alice and Bob cannot infer Jim's decision from the results of their $separate$ measurements.  For example, if---regardless of Jim's decision---Alice measures either $a$ or $a^\prime$, and obtains results $\pm 1$ with equal probability, and likewise Bob measures either $b$ or $b^\prime$, and obtains results $\pm 1$ with equal probability, then the unary condition is fulfilled.  The $binary$ condition states that if ${\hat a}$ is the spacetime event of Alice's measurements on her ensemble, ${\hat b}$ is the spacetime event of Bob's measurements on $his$ ensemble, and ${\hat j}$ is the spacetime of event of Jim pushing the button on his device, then the overlap of the forward light cones of ${\hat a}$ and ${\hat b}$ lies entirely within the forward light cone of ${\hat j}$.  (See Fig. 2.)  As shown in Ref. \cite{grunhaus1996jamming}, if jamming obeys the unary and binary conditions, then it is consistent with relativistic causality even though ${\hat a}$ and ${\hat b}$ may be {\it earlier} in time than ${\hat j}$.
\begin{figure}
\centerline{
\includegraphics*[width=150mm]{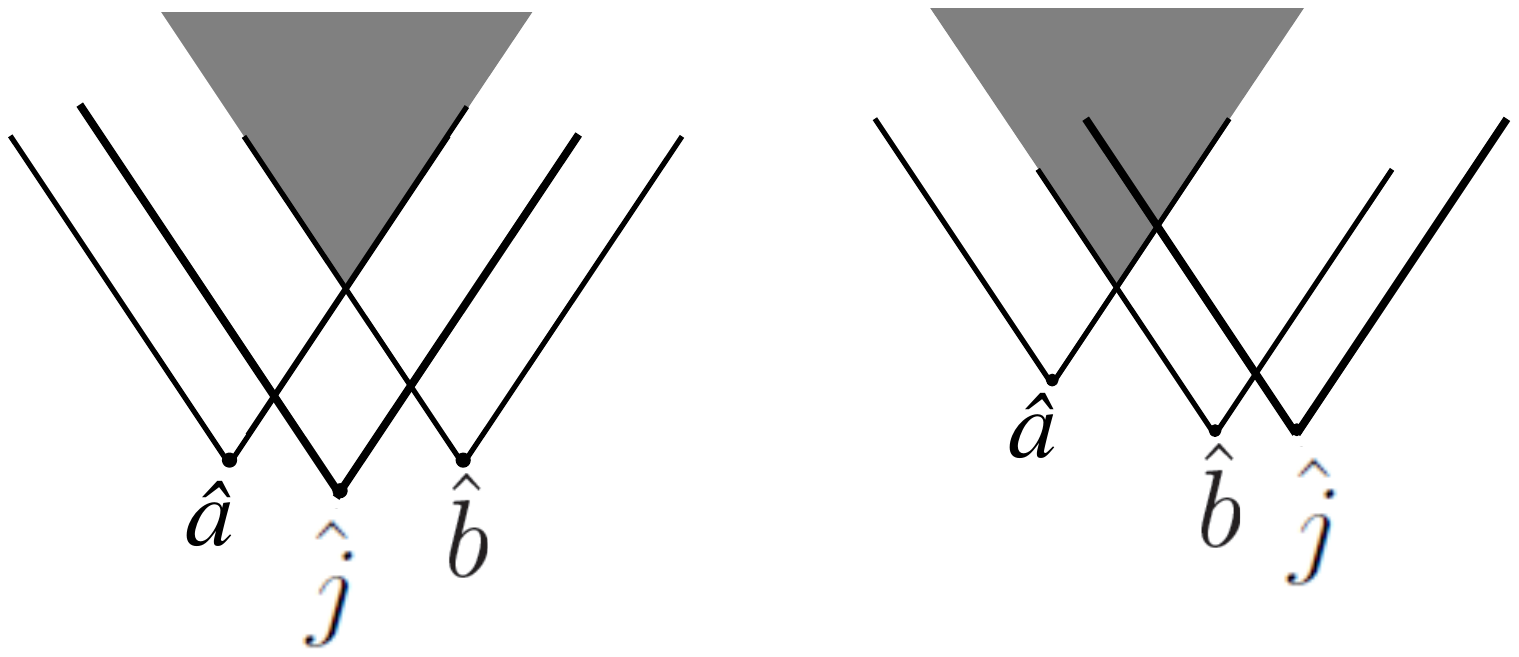}}
\caption[]{The overlap of the future light cones of ${\hat a}$ and ${\hat b}$ either (a) lies or (b) does not lie entirely within the future light cone of ${\hat j}$.}
\label{Fig2}
\end{figure}

We return now to the GHZ correlations of Eq. (\ref{GHZcorr}) and show that they permit jamming \cite{rohrlich2012three}. Suppose Alice, Bob and Jim share an ensemble of particle triplets in the GHZ state. If Jim consistently measures $\sigma_z^J$, he disentangles Alice's particles from Bob's, regardless of the outcomes he gets. If he measures $\sigma_x^J$, Alice's particles remain entangled with Bob's particles, and their spins are correlated. For example, $\sigma_x^A$ and $\sigma_x^B$ are perfectly correlated or perfectly anticorrelated, depending on Jim's outcome. If the information regarding Jim's outcomes is delivered to Alice and Bob, they can bin their $\sigma_x$ measurements in two ensembles corresponding to Jim's outcomes $\pm 1$. They will find that their results, within each ensemble, are perfectly (anti-)correlated in the case that Jim had chosen to measure $\sigma_x^J$, or uncorrelated in case he had measured $\sigma_z^J$.

This realization of jamming satisfies the unary condition because, regardless of Jim's decision, Alice's measurements of $\sigma_x^A$ average to zero, and likewise for Bob's measurements of $\sigma_x^B$.  It fulfills the binary condition because Jim must report to Alice and Bob the results of his measurements of $\sigma_z^J$ or $\sigma_x^J$ for them to determine, from the results of {\it their} measurements, whether their pairs were entangled or not.  Now, Alice and Bob can make their determination {\it only} in the overlap of the future light cones of ${\hat a}$ and ${\hat b}$, which must lie in the future light cone of ${\hat j}$ for them to receive Jim's input.  Thus jamming via GHZ triplets is consistent with relativistic causality. Nevertheless, Jim's decision, whether to leave the pairs shared by Alice and Bob in entangled or product states, can take place even {\it later} than ${\hat a}$ and ${\hat b}$, and even at a timelike separation from both measurements ${\hat a}$ and ${\hat b}$. (See Fig. 3.) Even then, it is only in the forward light cone of ${\hat j}$ that Alice and Bob can combine their data and determine whether Jim jammed their measurements.

\begin{figure}
\centerline{
\includegraphics*[width=140mm]{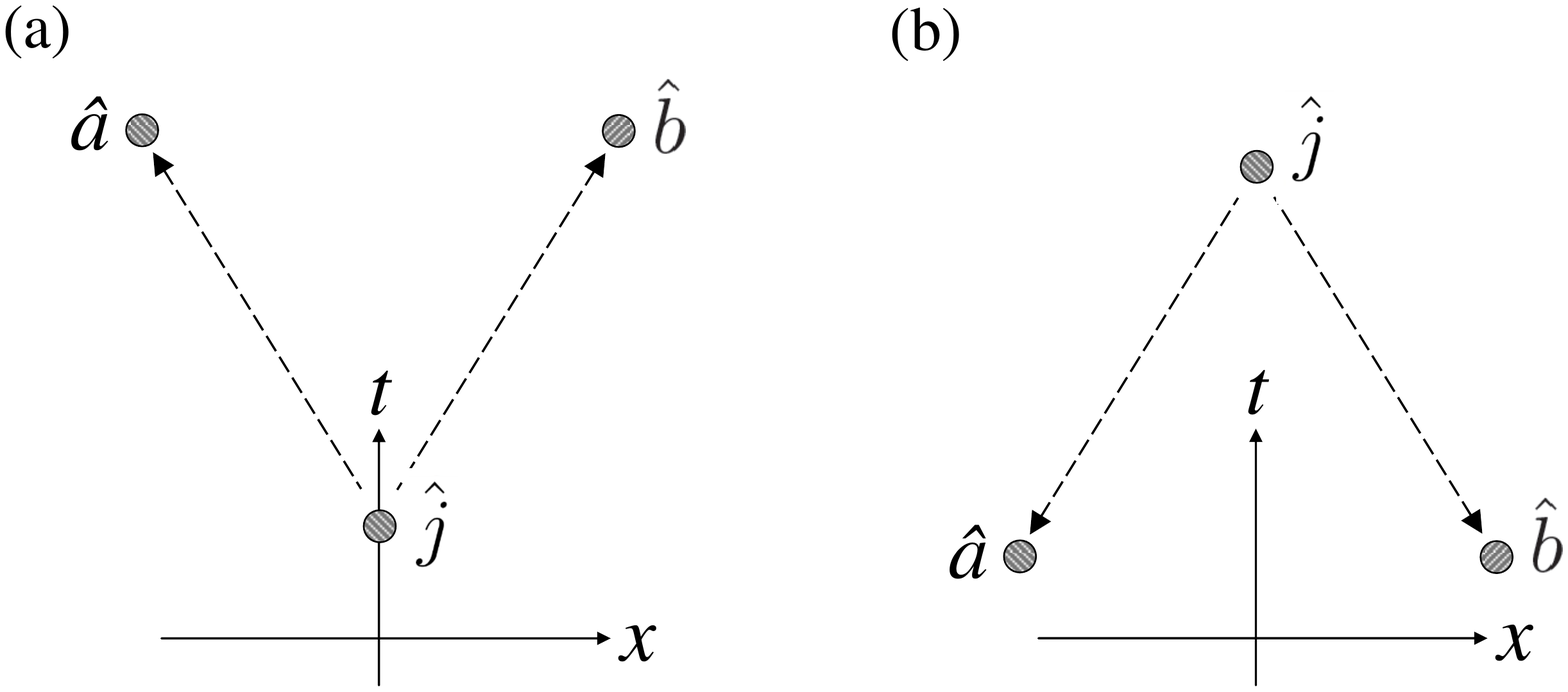}}
\caption[]{Configurations in which Jim can (a) causally and (b) retrocausally put pairs of particles shared by Alice and Bob in product or entangled states, as he chooses.  The dashed arrows connect cause with effect.}
\label{Fig3ab}
\end{figure}

So what makes PR-box correlations different from GHZ correlations, such that the former violate relativistic causality (in the classical limit) while the latter do not?  We might have replied, ``PR-box correlations are retrocausal whereas GHZ correlations are not."  But we have just seen that this distinction fails.  So let us return to our comparison, in the first section, of PR-box correlations and bipartite quantum correlations.  We noted that even quantum correlations that violate the Bell-CHSH inequality maximally are not strong enough to permit signalling.  Are GHZ correlations, which like PR-box correlations can be 0 or 1, strong enough?  No!  They are indeed stronger, but their strength dissipates over the {\it two} stages Alice and Bob require in attempting to receive Jim's signal.  Relativistic causality in the classical limit is a subtle, but effective, constraint on quantum mechanics.

We introduced this work by stating that three axioms with clear physical meaning, namely nonlocality, relativistic causality, and the existence of a classical limit, might be sufficient for deriving quantum mechanics, or at least an important part of the theory.  We can consider reducing these three axioms to two simply by eliminating nonlocality as an axiom.  Indeed, axioms in physical theories are, in general, constraints.  The constraint of locality could be an axiom, but absence of this constraint need not be an axiom.  And it seems from our work that quantum mechanics is just as nonlocal as it can be without violating relativistic causality.  The retrocausality we have seen in jamming via GHZ correlations suggests that also retrocausality, like nonlocality, can appear wherever it is not forbidden by relativistic
\nobreak
causality.
\goodbreak

\begin{acknowledgments}
This publication was made possible through the support of grants from the John Templeton Foundation (Project ID 43297), from the Israel Science Foundation (grant no. 1190/13).  The opinions expressed in this publication are those of the authors and do not necessarily reflect the views of either of these supporting foundations.

\end{acknowledgments}

\end{document}